\documentclass[acmsmall,screen,nonacm]{acmart}
\AtBeginDocument{%
  }

\usepackage{tikz}
\usetikzlibrary{positioning, fit, backgrounds, arrows.meta, shapes.geometric}
\usepackage{amsmath}

\usepackage{listings}
\usepackage{xcolor}
\usepackage{subcaption}
\usepackage{adjustbox}   

\definecolor{codegreen}{rgb}{0,0.6,0}
\definecolor{codepurple}{rgb}{0.58,0,0.82}
\definecolor{codeblue}{rgb}{0,0,0.8}

\newcommand{\coolname}{\textsc{Elevator}}

\begin{document}

\title{Deterministic Fully-Static Whole-Binary Translation without Heuristics}

\author{Hongyu Chen}
\email{hongyc4@uci.edu}
\affiliation{%
  \institution{University of California, Irvine}
  \city{Irvine}
  \state{California}
  \country{USA}
}

\author{James McGowan}
\email{jwmcgowa@uci.edu}
\affiliation{%
  \institution{University of California, Irvine}
  \city{Irvine}
  \state{California}
  \country{USA}
}

\author{Michael Franz}
\email{franz@uci.edu}
\affiliation{%
  \institution{University of California, Irvine}
  \city{Irvine}
  \state{California}
  \country{USA}
}

\renewcommand{\shortauthors}{Chen, McGowan, and Franz}


\begin{abstract}

We present \coolname{}, the first binary translator capable of statically translating entire x86-64 binary executables to AArch64 without using debug information, source code, or any assumptions about code patterns or layouts within the original binary. Unlike existing binary translation systems, which rely on heuristics or runtime fallback mechanisms to recover from erroneous code-versus-data decoding decisions, \coolname{} considers all possible interpretations of every byte in the original executable and creates a separate translation for each feasible one ahead of time. For example, any byte might be interpreted as data, as part of an opcode, or as part of an argument to an opcode. We generate separate control flow paths for all of these interpretations, pruning only those that lead to exceptional program termination.

For each such control flow path, a translation is generated by composing code ``tiles'' that have been automatically derived from a high-level language description of the individual instructions in the source instruction set architecture. This leads to a nimble binary translation framework.

Our approach is deterministic and produces complete, self-contained output binaries. Unlike previous solutions, it requires no runtime component in the trusted code base. \coolname{} offers a different cost/benefit profile than previous solutions, with the principal cost being a substantial code size expansion. The main benefit in return is that \coolname{}'s output is the actual code that will run, so it can be tested, validated, certified, and/or cryptographically signed before deployment, reducing risk compared to emulators or just-in-time (JIT) compilers.

We demonstrate \coolname{}'s effectiveness on a diverse corpus of real-world binaries, including the entire SPECint 2006 suite, showing that static full-program binary translation can be made both reliable and practical. Our method achieves performance on par or better than QEMU's user-mode emulation with JIT acceleration.

\end{abstract}




\settopmatter{printacmref=false}
\setcopyright{none}
\renewcommand\footnotetextcopyrightpermission[1]{}
\pagestyle{plain}

\maketitle

\section{Introduction}

Hardware transitions from one instruction set architecture (ISA) to another often come with a need to bring along legacy software to the new platform.
Often enough, such legacy software transition cannot be achieved fully mechanically ``simply by recompiling'' surviving source code.
To further complicate matters, when legacy code has been validated or certified, it is typically not the source code that is certified, but a specific well-tested ``authoritative binary executable.''

Recreating this exact ``authoritative binary'' bit-by-bit from source code at a later date is often impossible. Even if there is surviving source code and we manage to get it to compile and build, recreating the identical code is likely to require the exact same version of the compiler, linker, and possibly other parts of the build system that were used in the creation of the original binary.
As a consequence, recreating legacy programs from old source code may be risky because we can't be sure we are faithfully replicating the functionality of the ``authoritative binary.'' Instead, any surviving archived source code might refer to a subtly different version.
There are also reported cases, e.g.~\cite{microsoft2017patch}, in which a manufacturer fixed a software error by manually and skillfully applying a ``patch'' directly to the binary, bypassing source code altogether.
Utilizing the archived source code version might hence bring back unknown programming errors that had already been fixed in the currently running binary. Other than by exhaustive testing, there is typically no easy way of determining whether
any software recreated from source code at a later date
is actually the intended software version.

Instead of using source code, an alternative approach starts with the existing binary. As we expand on below in our related work section, previous solutions to working with binary code directly have employed combinations of emulation and static and dynamic translation. Common to all of these previous solutions is that they do not fully statically translate entire binary programs from one ISA to another, but that they all require additional system-level components that need to execute alongside the translated program.
These additional runtime components must therefore be part of the trusted code base and are implicitly included in all testing.
Ascertaining overall reliability is made more difficult by the possibility that the dynamic behavior of such systems could lead to different results based on the ordering of specific tests or inputs.

In contrast, \coolname{} makes the following key contributions:
\begin{itemize}
  \item \textbf{A fully static, deterministic, heuristics-free cross-ISA binary-to-binary translator.} \coolname{} is, to our
  knowledge, the first cross-ISA binary-to-binary translator from x86-64\footnote{We have implemented our system for x86-64 input binaries, but for better readability we mostly abbreviate this as
  ``x64'' throughout the text.} to AArch64 that is entirely static, fully deterministic, and heuristics-free. It makes no assumptions about the code layout of the input or the toolchain that produced it. Running the same input binary through \coolname{} twice yields the exact same output bit sequence. Once translation completes, the resulting AArch64 binary is a stand-alone executable that
  requires no runtime translation support and can be tested and certified in its own right.

  \item \textbf{A lightweight, LLVM-backed code generator.} Our code
  generator for AArch64 is built on a lightweight mechanism that
  leverages LLVM's mature compiler infrastructure to synthesize
  cross-ISA translations automatically, rather than being hand-written
  per instruction. This substantially reduces the engineering effort
  required to bring up a new back-end, and the same approach extends
  directly to other target architectures.
\end{itemize}

We have constructed a full-scale implementation prototype and demonstrate its effectiveness on a comprehensive test corpus. Our evaluation includes the entire SPECint 2006 benchmarking suite (as a proxy for real-world legacy binaries) and a small number of hand-crafted binaries (as exemplified by Listings~\ref{lst:overlap} and~\ref{lst:weird-branch}) designed to expose the limitations of existing fully static approaches.

We believe that our approach creates a useful new capability that is truly different and complementary to existing methods. There are situations in which our technique is likely to be superior to existing solutions, for example when certain processor models suddenly aren't available for political reasons or supply-chain issues. Using static translation of an existing binary can provide a rapid temporary ``stop-gap'' cross-ISA portability in this situation while preserving the ability to exhaustively test the resulting output code before it is deployed. This is less risky than using emulators or just-in-time compilers.

We have no commercial interests in this research and pledge to open-source everything at the end of our research project.

\section{Background and Related Work}
\label{sec:related}

There has long been an interest in modifying existing binaries without access to source code. The most general form of this is called \emph{binary rewriting} and aims to enable the application of various program transformations  to a program’s binary form.  The eventual goal of such transformations could be instrumentation, security hardening, optimization, or deobfuscation~\cite{Wenzl2019}. In the past few years especially, there has been a renewed interest in binary rewriting for a wide range of downstream applications that include security~\cite{microsoft-binary-patch}, optimization~\cite{2019Panchenko}, and code debloating~\cite{2019Razor, 2019Nibbler}. The term \emph{binary recompilation} is now frequently used to describe whole-program rewriting techniques that operate by first ``lifting’’ a program to an intermediate representation and then ``lowering’’ it back into a machine-executable form. The system we describe in this paper performs binary \textbf{cross}-(re)compilation from one ISA to another.

\emph{Static rewriters} operate on a binary file without executing it.
Static techniques range from being \textit{direct} to \textit{minimally-invasive} ~\cite{duck2020binary} to \textit{full-translation}~\cite{Wenzl2019}.
Direct and minimally-invasive schemes target specific tasks such as diverting control flow, inserting trampolines, or performing instruction-level modifications.
Full-translation techniques, on the other hand, usually translate programs to specialized intermediate representations (IRs) and eventually \textit{reassemble} a new binary.
IRs used by such rewriters aim to faithfully represent the original program semantics.
Examples include VEX IR~\cite{vex-valgrind}, Binary Analysis Platform’s (BAP) IL~\cite{brumley2011bap}, and REIL~\cite{dullien2009reil}.
Crucially, full-translation techniques use the expressivity of such powerful IRs to recover higher-level constructs such as control flow, basic blocks, and functions, which enables them to apply complex program-wide analyses and transformations.

\emph{Dynamic rewriters} transform a program \textit{during} program execution.
This is achieved by using an instrumentation engine such as PIN~\cite{luk2005pin} or DynamoRIO~\cite{bruening2003infrastructure} that inserts fine-grained hooks during native execution,
 or by running the binary inside a virtual environment such as QEMU~\cite{bellard2005qemu} or Valgrind~\cite{nethercote2007valgrind}.
Compared to static rewriters, dynamic approaches can perform much more precise and fine-grained modifications as they can observe control flow and program state at runtime.
However, modifications performed by such rewriters only persist for the duration of the current execution run.

Existing static binary recompilers have been surprisingly limited in their capabilities until quite recently; most of the published recompilation frameworks are not even able to recompile all of the constituent programs of some of the most basic standard benchmarking suites such as SPECint 2006. This is fundamentally due to the fact that ``lifting’’ is a hard problem:  Horspool and Marovac \cite{horspoolAndMarovac1980detranslation} showed as far back as 1980 that the general problem of ``detranslating’’ (decompiling/disassembling) a binary executable requires being able to differentiate with certainty between code and data, which for most computer architectures is equivalent to the Halting Problem~\cite{turing1967halting} and is hence unsolvable in general. Our approach overcomes this problem by translating each byte of the executable under all possible interpretations separately, and hence not having to make this determination at all.

Previous static binary lifters \cite{anand2013secondwrite, Dinaburg2014, Yadavalli2019, difederico2017revng, WilliamsKing2020, 2019Panchenko, dinesh2020retrowrite} have attempted to approximate the differentiation between code and data by using imprecise heuristics, which becomes a problem especially when trying to predict targets of indirect control flow transfers~\cite{pang2021sok}. For example, LLBT ~\cite{translator-for-arm1, translator-for-arm2} performs static translation of ARM binaries by lifting ARM instructions to LLVM IR and recompiling to the target architecture. However, like many static binary rewriters, LLBT relies on heuristics to detect potential indirect branch targets, rendering it vulnerable to incorrect translation when processing obfuscated or manually crafted indirect branching code. Additionally, LLBT makes several other assumptions about the input binaries during code identification, which help to shrink the size of their address mapping table and make the output binaries suitable for embedded devices where binary size is a serious concern.

The use of such imprecise heuristics is the main reason why all of the existing recompilers based on static binary lifting have problems handling even relatively simple benchmark programs: even a good heuristic will fail on some inputs, while correct lifting of an entire binary requires that the heuristic gets every single code-versus-data decision right. Hence, the larger a binary becomes, the higher the chances that at least one heuristics-based decision will come up wrong somewhere.

Conversely, dynamic approaches follow the flow of instructions as they are actually executed~\cite{altinay2020binrec}. They are thereby able to handle not only precise instruction recovery but also indirect control flows, by design; after all, the processor must also be able to follow and correctly decode the instruction stream. However, dynamic methods can only lift instructions that are reached during concrete executions of the program. Hence, a strictly dynamic lifter may have an incomplete view of the control flow graph contained in a binary program, and the parts of the binary that correspond to ``unseen’’ parts will be omitted from the lifted code. As a consequence, any output binary generated from the lifted code will have to be able to deal with situations in which a dynamically computed branch suddenly jumps into ``terra incognita,’’ i.e., a previously unseen piece of code that was not covered during lifting (and that may have been mistaken as ``data’’). Static lifters also have this problem when dealing with control flow targets that are reachable only via computed branches, some of which may evade regular static control-flow analysis.

To make matters even more complicated, ISAs such as x64 that have variable-length instructions make it possible to nest instruction sequences within each other. A branch terminating in the middle of a multi-word instruction will result in the operands of the original instructions being decoded as instructions in their own right; ``return-oriented programming’’ (ROP) attacks~\cite{schacham07geometry} frequently make use of this fact, but this strategy may also be used for code obfuscation. Hence, in order to statically cross-compile an entire binary, one needs to not only correctly take into account differentiation between code and data, but also all possible valid instruction sequences that may be embedded, at an offset, within other valid instruction sequences.

Current best practice is to combine static and dynamic approaches to handle precisely these corner cases. For example, commercial companies have employed such hybrid approaches when transitioning between ISAs, combining an interpreter for the old ISA along with a dynamic code generator for the new one.  Apple employed a system called Rosetta II in their transition from x64 to AArch64 on Mac systems ~\cite{rosetta_sunsetted}, a hybrid dynamic binary translation approach that performs some instruction translation ahead-of-time while translating others dynamically upon first discovery ~\cite{prj_champ_rosetta2}. Similarly, Microsoft has deployed a system called Prism ~\cite{prism} to support their ``Windows on Arm (WoA)’’~\cite{windowsonarm} initiative, combining ahead-of-time translation with dynamic components to handle undiscovered code and edge cases.

Among recent academic contributions, WYTIWYG~\cite{parzefall2024you} and Polynima~\cite{deshpande2024polynima} perform static binary lifting along control flow paths that have previously been identified by dynamic profiling; they also rely on fall-back mechanisms that collect additional control-flow information dynamically whenever branches terminate at previously unseen target addresses.

In contrast to all of these prior works, we present what we believe is the first general, fully static, heuristics-free binary cross-compiler that scales even to large programs. The key to our approach is that instead of ever making a determination as to whether any byte in the original program binary should be interpreted as code or data, as an instruction word or an argument to such an instruction, we create a separate control flow path under every feasible interpretation of that byte. This is an application of the concept of \emph{superset disassembly}~\cite{bauman2018superset}, which was first described in the context of binary rewriting, but which to the best of our knowledge has not been applied to static recompilation, let alone cross-compilation from one ISA to another. Our work here demonstrates that the approach can be made both robust and practical, while our measurements reveal some of the consequences of trading decoding precision for code expansion in this manner.

\section{System Overview}
\label{sec:overview}
\coolname{} operates on the principle of complete x64 state preservation within the translated AArch64 code. The system employs a one-to-one mapping between x64 and AArch64 registers, emulating the state of each x64 register within a corresponding AArch64 register. The x64 stack is emulated directly on the
AArch64 stack, allowing the operating system to handle normal stack expansion as the program executes. Rather than analyzing the application binary interface (ABI) of the input x64 binary, \coolname{} performs ABI translation only when execution transitions to and from external code; at those transition
points, the standard rules of the x64 System~V ABI and the AArch64 Procedure Call Standard regarding argument placement in registers and on the stack can be applied directly. The combination of complete state preservation and one-to-one register correspondence is what enables \emph{independent} translation of instructions: each x64 instruction can be translated with no knowledge of what runs before or after it, because the live state it reads and writes sits in the same dedicated AArch64 registers at every program point. Instruction-level isolation in turn allows us to represent the input as a superset control-flow graph (CFG), derived byte by byte from the original binary, and to translate each candidate x64 instruction in the graph into AArch64.

Translating an individual candidate instruction under this discipline is mechanical. The hard part of the pipeline is constructing the superset CFG itself. As mentioned in Section~\ref{sec:related}, distinguishing code from data in a binary is in general equivalent to the Halting Problem~\cite{turing1967halting} and therefore unsolvable; every static translator that attempts to commit, at each byte of the input, to a single interpretation is forced to rely on heuristics that will be wrong on some inputs. Listings~\ref{lst:overlap} and~\ref{lst:weird-branch} exhibit two pathological but entirely legal x64 programs that illustrate some of these issues.

\noindent
\begin{minipage}[t]{0.48\textwidth}
\begin{lstlisting}[language={[x86masm]Assembler},
  caption={Overlapping instruction example.},
  label={lst:overlap},
  basicstyle=\ttfamily\footnotesize,
  columns=fullflexible]
OverlappingInstruction:
    xor   eax, eax  ; zero result
    mov   al, 0xC2  ; load 0xC2
    test  rdi, rdi  ; test input
    jz    ReturnC2  ; exit if zero
ReturnC3:
    .byte 0xB0      ; MOV AL,0xC3
ReturnC2:           ;   byte 1
    .byte 0xC3      ;   byte 2
    ret
\end{lstlisting}
\end{minipage}%
\hfill
\begin{minipage}[t]{0.48\textwidth}
\begin{lstlisting}[language={[x86masm]Assembler},
  caption={Computed indirect branch.},
  label={lst:weird-branch},
  basicstyle=\ttfamily\footnotesize,
  columns=fullflexible]
WeirdIndirectBranch:
    and   rdi, 3    ; mask [0..3]
    shl   rdi, 1    ; offset * 2
    xor   eax, eax  ; set to zero
    call  Label     ; return address = base of table
    inc   eax       ; eax = 4
    inc   eax       ; eax = 3
    inc   eax       ; eax = 2
    inc   eax       ; eax = 1
    ret
Label:
    pop   rsi       ; pop address of table
    add   rsi, rdi  ; add offset
    jmp   rsi       ; computed jmp
\end{lstlisting}
\end{minipage}

Listing~\ref{lst:overlap} illustrates the nested-instruction phenomenon. Starting a decode at the \texttt{.byte 0xB0} yields \texttt{MOV AL, 0xC3} followed by \texttt{RET}, whereas starting one byte later at \texttt{ReturnC2} yields just \texttt{RET}. Both decodes are reachable from the preceding \texttt{jz}, and any translator that commits to a single interpretation of these two bytes will silently miss one of them. Listing~\ref{lst:weird-branch} illustrates a computed indirect branch: the \texttt{call} instruction captures the table's base address which is recovered by \texttt{pop rsi}, to which an input dependent offset is added to construct the target of \texttt{jmp rsi}, so the branch may land at any of four \texttt{inc eax} instructions that lie two bytes apart in the encoded stream. A translator that rewrites only statically resolvable jump targets has nowhere to land this branch.

\coolname{} sidesteps the code-versus-data determination altogether through an application of superset disassembly~\cite{bauman2018superset}: we simultaneously interpret every executable byte offset in the original binary as (i) data and (ii) the start of a potential instruction sequence beginning at that offset, and we build the superset control flow graph from every one of the resulting candidate decodes. Every potential target of indirect jumps, callbacks, or other runtime dispatch mechanisms that cannot be statically analyzed therefore has a corresponding landing point in the rewritten binary. These targets are resolved at runtime through a lookup table from original instruction addresses to translated code addresses that we embed in the final binary.


\coolname{}'s translation process falls into two distinct stages. The first, executed once offline and independent of any input binary, constructs a reusable \emph{tile bank}: a set of precompiled AArch64 byte sequences, one for every concrete combination of an x64 instruction and its operand registers. Each compiled sequence reads and writes the AArch64 registers that hold the emulated x64 state directly. The second, executed on every input binary, performs superset disassembly, selects the appropriate tile from the bank for each candidate x64 instruction it has discovered, concatenates the selected tiles into the body of an AArch64 object file, embeds the address-lookup table alongside, and links the object against the \coolname{} runtime driver to produce the final stand-alone executable. Section~\ref{sec:translate_cfg} develops both stages, beginning with the offline construction of the tile bank.

\begin{figure}[h]
\centering
\begin{tikzpicture}[
  font=\footnotesize, >=stealth, thick, node distance=8mm,
  box/.style={
    draw, rounded corners=1.5pt, align=center,
    minimum height=9mm, minimum width=24mm,
    inner sep=2pt, fill=white
  },
  artifact/.style={box, fill=gray!10},
  flow/.style={->, semithick}
]

\node[artifact]              (binary) {x86-64 binary};
\node[box, right=of binary]  (elev)   {\coolname{}};
\node[artifact, right=of elev] (exe)  {AArch64 executable};

\draw[flow] (binary) -- (elev);
\draw[flow] (elev)   -- (exe);

\node[artifact, above=8mm of elev, align=center, minimum height=6mm] (tb)
  {Tile bank\\{\tiny (built offline)}};

\draw[flow] (tb) -- (elev);

\end{tikzpicture}
\caption{\coolname{} system overview. For each input x86-64 binary,
\coolname{} consults a reusable tile bank, built once offline from
hand-written C tiles compiled through LLVM with a custom calling
convention, and emits a stand-alone AArch64 executable.}
\label{fig:elevator-overview}
\end{figure}
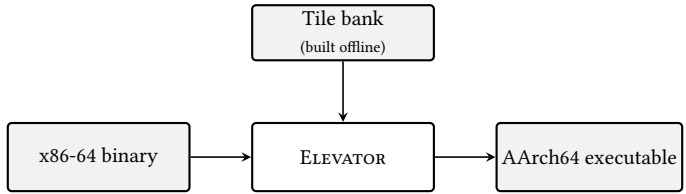

\section{Translating the CFG}
\label{sec:translate_cfg}
\coolname{} separates translation into three stages. An offline stage (Section~\ref{sec:tile-bank}) expresses x64 instruction semantics as C functions, specializes them per operand combination under a fixed x64-to-AArch64 register mapping, and compiles the whole set through a modified LLVM~20 into a reusable \emph{tile bank}. A per-binary stage (Section~\ref{sec:rewriting}) then rewrites an input x64 binary by looking each candidate instruction's tile up in the bank by name and stitching the retrieved AArch64 byte sequences together, with a small set of hand-crafted templates for the instruction categories that cannot be expressed as C tiles (control-flow transfers and ABI crossings). A final packaging stage (Section~\ref{sec:executable}) combines the translated code, the original x64 binary, an address-lookup table, and a runtime driver into the stand-alone AArch64 executable. Using an existing compiler to produce target-ISA code snippets that are then extracted and stitched together is an approach shared with template-based just-in-time compilers~\cite{piumarta1998optimizing, iliasov2003templates,
wimmer2013maxine, xu2021copypatch}; what is distinctive in our design is how tiles are specialized per x64 operand, how LLVM is configured so that the compiled tile code operates directly on emulated x64 state, and how the three stages cooperate so that the per-binary stage reduces to byte-level selection and concatenation.

\subsection{Offline: Building the Tile Bank (Performed only Once)}
\label{sec:tile-bank}

Writing a translation map from every x64 instruction to an equivalent AArch64 instruction sequence by hand, one assembly sequence at a time, is impractical. A single template such as the \texttt{ADD Reg8, Reg8} form already expands into 256 concrete register combinations, and the full x64 instruction set has many such templates across its register, memory-operand, and immediate addressing-mode variants; hand-writing an AArch64 encoding for each one would require dual expertise in both ISAs and an error-prone amount of effort. \coolname{} therefore does not write such a map directly. We express the semantics of each x64 instruction as a small C function, specialize it per concrete operand combination, and let LLVM compile the resulting set into AArch64. 

As a concrete example, Listing~\ref{lst:tile-template} shows the template tile for the x64 instruction \texttt{ADD Reg8, Reg8}, and Listing~\ref{lst:tile-specialized} shows the specialized tile that emerges for the particular instruction \texttt{ADD RCX, RDX}.

\noindent
\begin{minipage}[t]{0.48\textwidth}
\begin{lstlisting}[language=C,
  caption={Template tile for \texttt{ADD Reg8, Reg8}.},
  label={lst:tile-template},
  basicstyle=\ttfamily\scriptsize,
  columns=fullflexible]
uint64_t ADD8_R1_R1_R2(uint64_t R1, uint64_t R2) {
  return ((R1 + R2) & MASK8ULL) | (R1 & ~MASK8ULL);
}
\end{lstlisting}
\end{minipage}%
\hfill
\begin{minipage}[t]{0.48\textwidth}
\begin{lstlisting}[language=C,
  caption={Specialized tile for \texttt{ADD RCX, RDX}.},
  label={lst:tile-specialized},
  basicstyle=\ttfamily\scriptsize,
  columns=fullflexible]
__attribute__((aarch64_custom_reg("X3: X3, X2")))
uint64_t ADD8_RCX_RCX_RDX(uint64_t R1, uint64_t R2) {
  return ((R1 + R2) & MASK8ULL) | (R1 & ~MASK8ULL);
}
\end{lstlisting}
\end{minipage}

The template on the left returns the new value of the destination register: the lower eight bits are updated with the 8-bit sum while the upper 56 bits remain unmodified, matching the x64 semantics for partial-register writes. The x64 \texttt{ADD Reg8, Reg8} instruction also affects the \texttt{RFLAGS} register, modifying the Carry, Parity, Auxiliary Carry, Zero, Sign, and Overflow flags; since a C function is constrained to return a single value, we capture the flag updates in a separate, dedicated flag tile that runs alongside the arithmetic one. A single x64 instruction may therefore correspond to one or several tiles, which we concatenate back-to-back at emission time to recover the full semantics.

Two things change between the template on the left and the specialized form on the right. The function name has been rewritten from \texttt{ADD8\_R1\_R1\_R2} to \texttt{ADD8\_RCX\_RCX\_RDX}, pinning the template's positional register arguments to the concrete x64 operands \texttt{RCX} and \texttt{RDX}. An attribute \texttt{aarch64\_custom\_reg} has also been attached, declaring the AArch64 registers in which LLVM is to place the return value and each argument: in this example the return value and first argument both bind to \texttt{X3}, and the second argument binds to \texttt{X2}, reflecting the mapping \texttt{RCX}\,$\mapsto$\,\texttt{X3} and \texttt{RDX}\,$\mapsto$\,\texttt{X2}. The body of the function, which operates on the local variables \texttt{R1} and \texttt{R2}, is unchanged; the attribute is what causes LLVM to read those locals out of \texttt{X3} and \texttt{X2} on entry and to write the result back into \texttt{X3} on exit. Everything else in this subsection is about how the template is turned into the specialized form on the right, and how that specialized form is compiled and packaged into the tile bank.

A fixed x64-to-AArch64 register mapping is realized by a custom LLVM calling convention applied per tile, chosen under three constraints:
\begin{enumerate}

  \item \textit{Volatility preservation.} An x64 register that is callee-saved under System~V maps to an AArch64 register callee-saved under AAPCS64, and symmetrically for caller-saved registers, so that calls into AAPCS64 libraries preserve emulated x64 state across the boundary.

  \item \textit{Argument-position alignment.} The register holding the $n$-th integer argument under System~V maps to the register holding the $n$-th integer argument under AAPCS64, so that a call from translated code to an AArch64 library restates its positional arguments rather than reshuffling them across the call.

  \item \textit{Minimality.} We consume as few AArch64 callee-saved registers as the first two constraints allow. AArch64 offers twelve such registers against x64's seven, and we keep the surplus free for future shadow state without perturbing the existing mapping.
      
\end{enumerate}
Beyond the general-purpose registers, x64's \texttt{RFLAGS} bits and XMM register file are held in dedicated AArch64 registers under the same one-to-one discipline, keeping the full emulated state resident in the register file. 

Producing the tile bank itself is mostly mechanical from here. A modified LLVM~20 honors the \texttt{aarch64\_custom\_reg} attribute on a per-function basis and reclassifies the AArch64 registers backing emulated x64 state as callee-saved inside the register allocator, so that neither argument placement nor scratch usage inside a tile can corrupt emulated state. A small source-to-source tool, \texttt{TileGen}, walks each C template and emits one specialized copy per admissible operand combination, synthesizing the attribute mechanically from the template's parameter positions using the register mapping above. Compiling the specialized file through the modified LLVM, with a short post-pass that makes tile bodies concatenation-safe, yields the tile bank: a map from tile name to AArch64 byte sequence, built once offline and consumed by the per-binary stage described next.

\subsection{Rewriting an x64 Binary (Performed Once for Each Binary)}
\label{sec:rewriting}

Given an input x64 binary, the per-binary stage performs superset disassembly and walks the resulting CFG. For each node, a formatter derives the tile name from the decoded instruction's opcode and operands, composing multiple names for instructions whose effects span several tiles.

While x64 imposes no restriction on stack pointer alignment, AArch64 requires strict 16-byte alignment when using the stack pointer in memory operands. Although we emulate the x64 stack on the AArch64 stack, directly mapping \texttt{RSP} to \texttt{SP} creates several complications. The absence of alignment requirements in x64 creates frequent violations in regular code patterns. For example, consecutive \texttt{PUSH} instructions in function prologues guarantee non-16-byte-aligned memory accesses, which would trigger exceptions on AArch64. We address this by having tiles access the stack through a separate register, \texttt{X25}, and only materializing \texttt{SP} in it when tiles actually require it. Additionally, since our tiles compiled by LLVM expect 16-byte \texttt{SP} alignment upon entry, we align \texttt{SP} down prior to executing any tile detected as allocating spill space, restoring it either from a saved register or \texttt{X25} depending on if the tile modified \texttt{RSP}.

We also implement a targeted optimization to eliminate unnecessary flag computation tiles, which we identify as being relatively expensive when compared to other tiles. If the flags are overwritten prior to a read in a post-dominating instruction, the flag computation part of the current node's tile can be removed.


When we encounter unsupported instructions, which currently consist principally of x64's AVX2 and later wide-vector extensions, we insert an interrupt instruction in place of a tile. This has no practical impact: superset disassembly inherently decodes numerous invalid or spurious instruction sequences at arbitrary byte offsets, but these occur on program paths that are never reached during normal control flow. Our evaluation across all of SPECint~2006 demonstrates that our supported set, comprising the full x86-64 integer ISA and the SSE subset exercised by SPECint, is sufficient to execute every benchmark. Furthermore, \coolname{} is designed to be an extensible framework, so that adding new tiles to support additional instructions is a straightforward process; however, the extra engineering work is highly unlikely to yield any additional scientific insights.

\subsubsection{Control-Flow Instructions}
\label{sec:control-flow}

Call, return, and branch instructions cannot be expressed as C tiles: their semantics depend on architectural decisions (return-address location, program counter, conditional-flags layout) that differ between x64 and AArch64, so a naive mapping such as x64 \texttt{CALL} onto AArch64 \texttt{BL} would break the emulated x64 stack. We hand-craft the translation for each category.

\textbf{Call.} Direct calls need no ABI translation: we push the original x64 return address (the x64 call site plus five) onto the emulated stack and branch to the translated tile of the callee. The stacked address stays in the original x64 address space; the branch target is its translated counterpart. Indirect calls, whose target is known only at runtime and may land either inside the translated binary or inside an external library, emit a bounds check against the embedded x64 binary and branch accordingly. In both cases the original x64 return address is pushed first. For internal targets, the x64-offset-to-tile table translates the target before an unconditional branch to the corresponding tile. For external targets, we install the address of a reverse ABI-translation gadget in \texttt{X30} (where the AArch64 library will return to), perform the exit ABI translation, and branch to the external target.

\textbf{Return.} Returns pop the 8-byte return address from the emulated stack and compare it against the embedded x64 binary's bounds. Internal returns translate the address through the lookup table and branch to the corresponding tile; external returns perform the return-side ABI translation before branching to the target.

\textbf{Branch.} We split branches by whether the target is statically resolvable (direct branches) or computed at runtime (indirect branches). Direct branches encode their target as an immediate offset relative to \texttt{RIP}, so the target is known at translation time. Unconditional \texttt{JMP Imm8/Imm32} maps to AArch64 \texttt{B} with the offset re-encoded from x64's 8 or 32 bits into AArch64's 26-bit range. Conditional branches translate to the AArch64 conditional branches that test the x64 flag bits held in \texttt{X14}. In both cases we emit the branch with a placeholder offset and defer address fixup to the linker, which patches the final offset once all tiles have been placed and inserts a veneer if the 26-bit range is exceeded.

\begin{figure}[!h]
    \centering
    \begin{subfigure}[b]{0.45\columnwidth}
    \centering
    \raisebox{1.5cm}{\includegraphics[width=\linewidth]{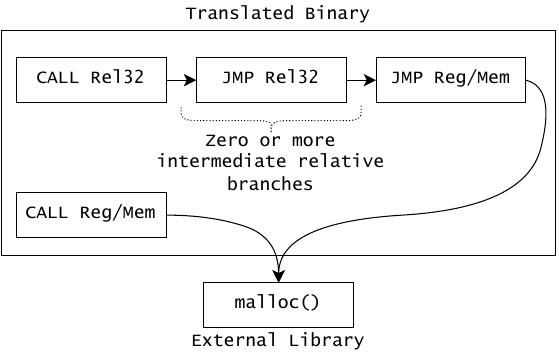}}
    \caption{Indirect Branch Leaving the Translated Binary.}
    \label{fig:indirect-jmp-chain}
    \end{subfigure}
    \hfill
    \begin{subfigure}[b]{0.5\columnwidth}
        \centering
        \includegraphics[width=\linewidth]{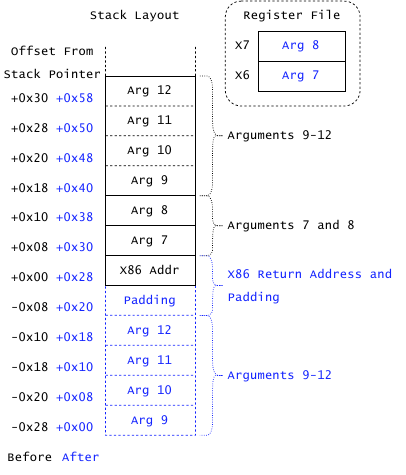}
        \caption{Exit ABI Translation with \texttt{n}=4 stack arguments.}
        \label{fig:exit-abi-translation}
    \end{subfigure}
    \caption{Indirect branch handling and ABI translation at exit boundaries.}
    \label{fig:indirect-and-abi}
\end{figure}

Indirect branches may target either the translated binary or an external library; the latter typically arises when a call to an external library at the end of a function is optimized into a tail call, as in Figure~\ref{fig:indirect-jmp-chain}. We emit the same bounds check used for indirect calls and returns, and perform exit ABI translation when the target is external. The indirect jump differs from the indirect call only in that the ABI-translation path assumes a preceding relative-call tile has already placed the return address at \texttt{[RSP]}.

\subsubsection{Crossing the x64/AArch64 ABI Boundary}
\label{sec:abi}

\coolname{} supports only dynamically linked binaries. This side-steps the need to translate architecture-specific instructions such as \texttt{CPUID}, which statically linked binaries inline directly but dynamically linked binaries delegate to libc. To facilitate this \coolname{} supports transitioning to and from the x64 and AArch64 Linux ABIs when interacting with dynamically linked libraries. There are four distinct cases where ABI translation is necessary as execution crosses between our emulated x64 environment and native AArch64 library code. The two aspects requiring translation are argument placement and return address location.
The System V x64 ABI (used by x64 Linux) designates six registers, \texttt{RDI}, \texttt{RSI}, \texttt{RDX}, \texttt{RCX}, \texttt{R8}, and \texttt{R9}, as argument registers, with additional arguments passed on the stack starting at \texttt{[RSP+8]}. The x64 \texttt{CALL} instruction stores the return address on the stack below any arguments at \texttt{[RSP]}. In contrast, the AArch64 Procedure Call Standard defines eight argument registers (X0-X7) with remaining arguments on the stack at \texttt{[SP]}, while also storing the return address in register \texttt{X30}.

\textbf{Calls to External Libraries.}
When a translated x64 call instruction targets an external library, we must change the argument layout to respect AArch64’s calling conventions. First, we subtract 8 from \texttt{SP} to realign the stack to a 16 byte boundary, leaving the x64 return address, which was already on the stack, at \texttt{[SP+0x8]}. We then load two values from stack positions \texttt{[SP+0x10]} and \texttt{[SP+0x18]} into registers \texttt{X6} and \texttt{X7}, giving AArch64 libraries access to potential arguments 7 and 8 that the translated x64 code would have placed on the stack if they exist. However, any potential remaining stack arguments are left starting in the wrong location, \texttt{[SP+0x20]}. Ideally we would have popped the x64 return address off the stack, as well as the values stored into \texttt{X6} and \texttt{X7}. Unfortunately, this is unsafe as we cannot be sure if the popped argument values are not instead caller spill space or part of a structure allocated on the caller’s stack. Instead, we leave the caller’s stack layout entirely untouched and allocate an additional \texttt{n×8} bytes of stack space.
In this new space, we copy in \texttt{n} potential 8-byte arguments from their current locations, starting at \texttt{[SP+0x20+n×8]}. This stack copy, starting at the new adjusted \texttt{SP}, now holds any potential stack arguments (arguments numbered 9 and above) that would be passed to the callee. Figure \ref{fig:exit-abi-translation} depicts the transformation applied to the stack and register file upon control flow leaving the translated binary, with black being the layout before, and blue being the updated layout after. The maximum number of arguments \coolname{} allows any function call in the input binary to pass is \texttt{n} stack arguments plus 6 register arguments, with \texttt{n} defaulting to 10. However, \texttt{n} is fully configurable and can easily be increased to support input binaries that call external library functions with more than 16 total arguments.
Finally, we store the address of a gadget in \texttt{X30} for the external library to return to.

\textbf{Returns from External Libraries.}
When control returns to the gadget whose address was stored in \texttt{X30} before the call to the external library, the previously copied stack arguments are cleaned up by adding \texttt{n×8} to the stack pointer. We then move the external library’s return value from \texttt{X0} into \texttt{X9 (RAX)}, where the emulated x64 code expects it. Finally, we pop the original x64 return address and its associated padding from the stack, translate the address, and branch to it, thereby resuming execution after the original \texttt{CALL}.

\textbf{Callbacks Into Translated Code.}
When native AArch64 code calls into our translated binary, we must convert from AArch64’s calling convention to x64’s. The emulated x64 code expects arguments 7 and 8 on the stack rather than in registers \texttt{X6} and \texttt{X7}. We push \texttt{X7} first, then \texttt{X6}, placing them at the stack positions where x64 would expect these arguments. If the callee does not actually expect a 7th and 8th argument, these pushed values will not have any effect. Finally, we push the return address, which the AArch64 branch-and-link instruction in the external library will have put in \texttt{X30}, onto the stack where the x64 return instructions will expect it.

\textbf{Callback Returns from Translated Code.}
When translated code returns from a callback to an external library, we reverse the entry process. The return address is popped off the stack, and the stack space allocated by pushing \texttt{X6} and \texttt{X7} is cleaned up by adding 0x10 to the stack pointer. Since our translated callback code will have put the return value in \texttt{X9(RAX)}, but the external library will expect it in \texttt{X0}, we move it from \texttt{X9} into \texttt{X0}.

\textbf{Architecture-dependent structures.} A few data structures have different layouts on x64 and AArch64 and must be translated in addition to the argument and return-address marshalling above. The most prevalent is \texttt{va\_arg}, whose layout reflects the ABI's argument-register count (six for System~V, eight for AAPCS64). We install intercept stubs on the affected library routines (\texttt{vsprintf} and \texttt{vsprintf\_chk}) that perform the extra x64-to-AAPCS64 \texttt{va\_arg} rewrite as control leaves the binary.

\subsection{The Translated Executable}
\label{sec:executable}
Once the superset CFG has been rewritten into an AArch64 code stream, the remaining work is to turn that stream, together with its auxiliary structures, into a stand-alone ELF executable. The translated binary embeds four components: the translated code itself, the original x64 binary preserved verbatim as read-only data, the address-lookup table that the runtime uses to resolve computed indirect branches, and a small runtime driver that installs memory protections and a signal handler at startup.

\subsubsection{Binary Layout}
\label{sec:binlayout}
\textbf{Translated Code.} The translated superset CFG is included in the new file by inserting the tiles into symbols for the linker to place. The system employs a greedy algorithm that traces static fall-through edges in the superset CFG forward and merges consecutive tiles into single symbols. Heuristics are used to identify likely call targets in the superset CFG, with the algorithm starting at these positions to merge likely executed tiles together. When a branching instruction is encountered, forward merging is stopped and restarted at the branch target(s). Labels to each tile within the merged symbols are preserved, as other tiles outside the symbol may still need to jump into them. When an instruction is encountered that has already been visited by the algorithm, meaning it already exists in another merged symbol, a branch to it is inserted and the linear merging stops. The symbols are passed to the linker (LLD), which places them into the final executable and adds relocations to the x64-offset-to-tile lookup table with the final tile positions.

\textbf{Original Binary.} We preserve the original x64 binary, embedding it in a new section within the translated binary. This embedded section contains a mapped view of the original binary with all segments expanded and positioned as they would appear if the ELF loader had loaded it into memory for execution. Initially, the embedded binary resides in a read-only section; however, during program startup, we parse the original program headers and apply the correct memory protections to each segment. Crucially, we intentionally avoid marking any originally executable sections as executable in the translated binary. None of the original code will be executed, and the fact that a signal will be raised if execution is attempted allows us to detect external calls into the binary, as will be explained later.

\textbf{Lookup Table.} The x64-offset-to-tile lookup table enables efficient translation between original x64 addresses and their corresponding translated code locations. The table is an array of 8 byte offsets into the translated binary, and is indexed into using offsets in the original binary.

\textbf{Driver.} The driver program is a small segment of C code linked into the translated binary that executes before the translated x64 code. The driver code takes on many of the responsibilities the original x64 loader would have performed, such as applying proper segment memory protections, processing PLT relocations, resolving external library symbols and populating GOT entries. It intercepts ABI incompatible library calls, such as \texttt{vsprintf}, redirecting them to separate functions to translate their architecture-specific structures.

\subsubsection{Signal Handler}
\label{sec:sighandler}
Range checks added to indirect call and jump instructions enable straightforward detection of when ABI transitions are necessary upon leaving the translated binary. However, detecting when control flow \emph{enters} the translated binary, which also necessitates ABI translation, is more difficult. \coolname{} solves this by installing signal handlers that catch attempts to execute code within the embedded original x64 binary, and subsequently perform ABI translation and redirect control flow to the corresponding tiles.

\textbf{Escaping Pointers.} Code pointers referencing our translated binary can escape to external libraries in only two ways: (1) function pointers passed as parameters to external libraries, and (2) exported functions. In both cases, these escaped pointers reference addresses within the original x64 binary, which in the translated binary will be non-executable. With this, we can be sure that every attempt to execute code within our binary, from an outside source, will result in attempted execution inside the embedded x64 binary. When this happens, a hardware exception is raised and execution is subsequently transferred to our signal handler where an ABI transition can be performed.

\section{Evaluation}

To evaluate \coolname{}, we first validate that \coolname{} preserves the functionality of the original x86-64 binaries. We then measure the static translation cost (Section~\ref{sec:eval:xlate-time}), followed by runtime performance compared against native AArch64 binaries and against two dynamic binary translators. The two translators cover the current state-of-art along different dimensions. QEMU~8.2.2~\cite{bellard2005qemu} in user-mode emulation is the de-facto reference for portable dynamic translation and the baseline used by nearly every prior binary-rewriting study on Linux. Box64~\cite{box64} is a mature, actively developed x86-64 to AArch64 dynamic translator originating in the gaming and Wine ecosystem, and represents the upper end of performance currently achievable for this ISA pair. We
emphasize up-front that Box64 is a production-quality engineering
artifact refined over many years by a large open-source community, with
extensive hand-tuned AArch64 dynarec code paths for common x86
instruction patterns; it is not directly comparable, on an
engineering-effort basis, to a research prototype such as \coolname{}. It is also not directly comparable on a structural basis. Box64 executes the x86-64 input through a runtime engine that JIT-compiles basic blocks on first encounter and that itself remains resident on every run, whereas \coolname{} produces a self-contained AArch64 ELF that replaces the input entirely at translation time. 

No dynamic translator of which we are aware can produce such an artifact, and the obstacle is fundamental to the dynamic approach rather than an engineering gap to close. Dynamic translators discover code by following the program counter at runtime, so computed branches, indirect calls, and any code introduced through \texttt{dlopen}, self-modification, or JIT compilation have no form that an ahead-of-time pipeline can consume. \coolname{}'s superset disassembly is what makes whole-program static translation tractable, and is the reason \coolname{} has a shippable artifact to produce at all.  We nevertheless include Box64 as an aspirational reference point: the gap between \coolname{} and Box64 quantifies the optimization headroom that more sophisticated code generation could eventually deliver on top of our static, superset-based approach. Finally, we analyze the binary size expansion introduced by the superset-based rewriting approach that underlies \coolname{}.

\textbf{Benchmarks.} Our evaluation uses the SPEC CPU2006 integer suite
(SPECint~2006). We focus exclusively on the integer benchmarks because
full translation of the SPECfp~2006 binaries would require additional
engineering to handle the SSE and x87 floating-point repertoire without
yielding proportional scientific insight into \coolname{}'s design. For
each benchmark we compile x86-64 inputs with \texttt{gcc}~13.3.0 at both
\texttt{-O2} and \texttt{-O3}. The
corresponding native AArch64 baselines are produced by recompiling the
same sources with \texttt{gcc}~13.3.0 at the matching optimization
level. All benchmarks use the reference (\texttt{ref}) input set.

We deliberately use SPECint~2006 rather than migrating to the newer SPEC~CPU~2017 suite, and we motivate the choice carefully because both suites now appear in the recent binary-translation literature. SPECint~2006 remains the benchmark of record for static binary rewriters and cross-ISA translators whose design \coolname{} most directly inherits from or is most directly comparable against, including Multiverse~\cite{bauman2018superset}, Egalito~\cite{WilliamsKing2020}, BinRec~\cite{altinay2020binrec}, HQEMU~\cite{hong2012hqemu}, and MAMBO-X64~\cite{MAMBO-X64}. Among these, Multiverse is the closest intellectual antecedent: its superset disassembly is the direct origin of \coolname{}'s superset-based rewriting pipeline. Retaining SPECint~2006 therefore places \coolname{} on directly comparable footing with the prior work that shares its technical foundation. We acknowledge that a separate line of recent work, including Biotite~\cite{biotite} and the LLVM-based DBT of Engelke~\cite{engelke2021llvmdbt}, reports on SPEC~CPU~2017. We regard the two benchmark choices as complementary rather than competing. x86-64 SPECint~2006 built with \texttt{gcc}~13.3.0 demonstrates \coolname{}'s ability to handle modern code generation while preserving direct comparability with previous work in the field of static binary translation.



\textbf{Experimental Setup.} All benchmark evaluations were performed
on an AArch64 server running Ubuntu~24.04.2~LTS with Linux kernel
6.8.0. The system is a GIGABYTE 1U Mt.\ Snow 1S chassis housing an
AMPERE Altra processor built on the Arm Neoverse-N1 microarchitecture,
with 64~single-threaded cores running at 3.0~GHz, 64~GB of DDR4-3200
ECC memory in a single NUMA node, and a 1~TB Samsung NVMe SSD. Each
core has 64~KB L1-d, 64~KB L1-i, and a private 1~MB L2 cache. We run
every (\textit{benchmark}, \textit{input}, \textit{execution mode})
combination three times under the
system's default frequency governor and report the median wall-clock
time measured by \texttt{/usr/bin/time~-v}. Hardware performance
counters—cycles, instructions, cache references, cache misses,
branches, and branch mispredictions—are collected on every run via the
Linux \texttt{perf~stat} interface; for each configuration we extract
the counters from the run whose wall-clock time equals the reported
median. 


\textbf{Special Cases and Modifications.} Benchmark 471.omnetpp
unconditionally throws a C++ exception to terminate itself; however,
\coolname{} does not yet support exception handling. We therefore
applied a minimal modification to a single function, replacing two
\texttt{throw} statements with equivalent \texttt{return} values. This
preserves identical control-flow semantics and does not affect the
benchmark's computational characteristics. Additionally, we crafted
several input test binaries that perform unconventional control-flow
tricks, including non-standard indirect branching and overlapping
instruction sequences. These were created to probe edge cases that only
an assumption-less superset-based CFG can handle, and thus to
demonstrate the effectiveness of superset disassembly as the rewriting
substrate underlying \coolname{}.

\subsection{Correctness}
\label{sec:correctness}
We validate \coolname{}’s correctness at individual instruction translation level and complete binary translation level.
We verify each tile by comparing its output against the corresponding x86 instruction executed on native x86 hardware. Each instruction has over 100,000 test inputs, combining randomly generated values with carefully selected edge cases that include boundary values, arithmetic overflow and underflow conditions, and flag combinations.

We also verify end-to-end correctness by comparing outputs of translated binaries against their original x64 counterparts. Across all SPECint 2006 benchmarks (at both O2 and O3 optimization levels) and our custom programs featuring unconventional control flow, all outputs match. This demonstrates that \coolname{} correctly preserves program semantics throughout the entire translation process.

\subsection{\coolname{} Translation Speed}
\label{sec:eval:xlate-time}

\begin{figure}[h]
  \centering
  \begin{minipage}[t]{0.49\linewidth}
    \centering
    \includegraphics[width=\linewidth]{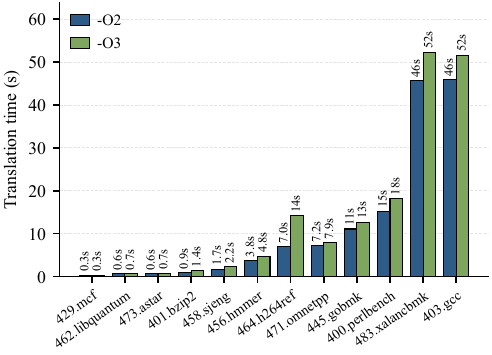}\\[2pt]
    {\small (a) Translation time.}
  \end{minipage}
  \hfill
  \begin{minipage}[t]{0.49\linewidth}
    \centering
    \includegraphics[width=\linewidth]{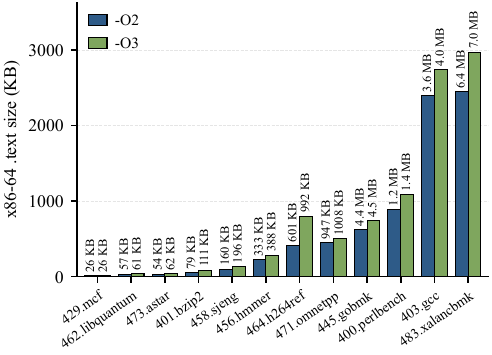}\\[2pt]
    {\small (b) x86-64 \texttt{.text} size; labels give total ELF size.}
  \end{minipage}
  \caption{\coolname{}'s translation time (a) tracks input
           \texttt{.text} size (b) at Pearson $r = 0.9993$; only
           403.gcc and 483.xalancbmk swap between the two orderings.}
  \label{fig:xlate-pair}
\end{figure}


Figure~\ref{fig:xlate-pair}(a) reports the translation time for each
SPECint~2006 benchmark at both \texttt{-O2} and \texttt{-O3}.
Translating the entire suite takes $140$~s at \texttt{-O2} and
$167$~s at \texttt{-O3}, and the per-benchmark cost is strongly
correlated with the size of the input x86-64 \texttt{.text} section
shown in Figure~\ref{fig:xlate-pair}(b) (Pearson $r = 0.9993$). The
annotations above each bar report the full ELF binary size, which
can be substantially larger than \texttt{.text} when a benchmark
bundles static data. \texttt{445.gobmk}, for instance, ships with a
$\sim$4.5\,MB pattern database. \coolname{} disassembles and
retranslates only \texttt{.text}, so it is that section's size that
governs the pipeline's cost.

\subsection{Runtime Performance}
\label{sec:runtime-performance}

\begin{figure}[t]
  \centering
  \includegraphics[width=\linewidth]{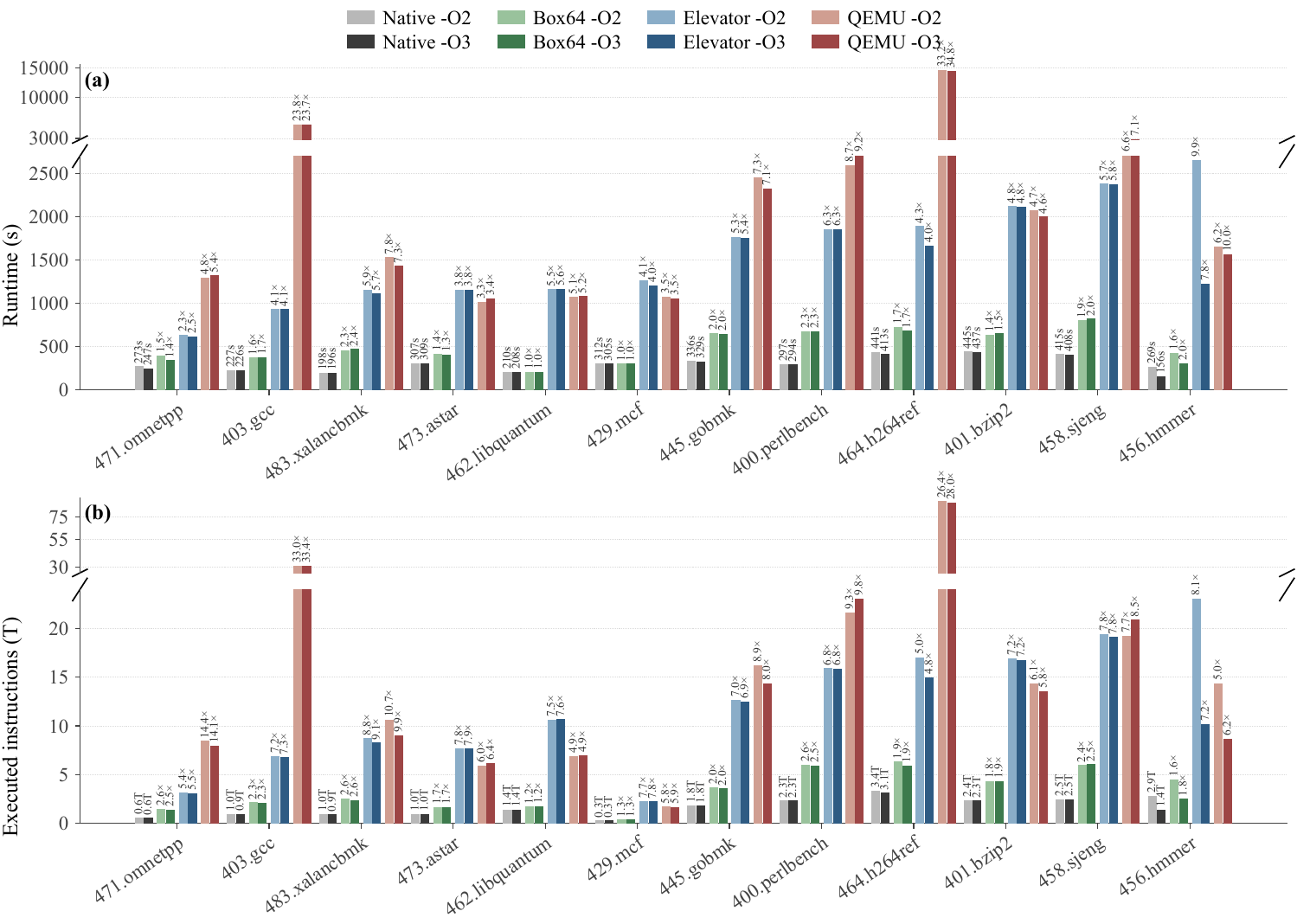}
  \caption{(a)~Wall-clock runtime and (b)~executed instructions
           on SPECint~2006 at \texttt{-O2}/\texttt{-O3} for
           native, Box64, \coolname{}, and QEMU. A broken
           y-axis compresses the \texttt{403.gcc} and
           \texttt{464.h264ref} QEMU outliers so the main data
           range stays at high resolution.}
  \label{fig:runtime-instructions}
\end{figure}

\begin{table}[h]
  \begin{minipage}[t]{0.49\linewidth}
    \centering
    \caption{Geometric-mean runtime slowdown vs.\ native
             AArch64 on SPECint~2006. Lower is better.}
    \label{tab:gmean-runtime}
    \begin{tabular}{lcc}
      \hline
      System      & \texttt{-O2}   & \texttt{-O3}   \\
      \hline
      Box64       & $1.58\times$   & $1.62\times$   \\
      \coolname{} & $4.88\times$   & $4.79\times$   \\
      QEMU        & $7.24\times$   & $7.69\times$   \\
      \hline
    \end{tabular}
  \end{minipage}\hfill
  \begin{minipage}[t]{0.49\linewidth}
    \centering
    \caption{Geometric-mean executed-instruction inflation
             vs.\ native AArch64 on SPECint~2006. Lower is
             better.}
    \label{tab:gmean-instructions}
    \begin{tabular}{lcc}
      \hline
      System      & \texttt{-O2}   & \texttt{-O3}   \\
      \hline
      Box64       & $1.94\times$   & $1.96\times$   \\
      \coolname{} & $7.12\times$   & $7.05\times$   \\
      QEMU        & $9.36\times$   & $9.56\times$   \\
      \hline
    \end{tabular}
  \end{minipage}
\end{table}

\subsubsection{Observed Runtime}
Figure~\ref{fig:runtime-instructions}(a) and
Table~\ref{tab:gmean-runtime} report wall-clock runtime on
SPECint~2006. \coolname{} slows execution down by a
geometric-mean factor of $4.88\times$ at \texttt{-O2} and
$4.79\times$ at \texttt{-O3} relative to the native AArch64
baseline, with per-benchmark slowdowns between
$2.34\times$ and $9.85\times$ at \texttt{-O2} and between
$2.50\times$ and $7.85\times$ at \texttt{-O3}. QEMU user-mode
is both slower on average and substantially more variable:
geometric means of $7.24\times$ and $7.69\times$, with
per-benchmark ranges of $3.33\times$--$33.20\times$ and
$3.43\times$--$34.81\times$. Box64, which we include as a
reference to the performance achievable by a mature production
dynamic translator, slows execution down by $1.58\times$
and $1.62\times$ on average. \coolname{} executes faster than
QEMU on 7 of 12 benchmarks at \texttt{-O2} and on 8 of 12 at
\texttt{-O3}. The rest of this section
decomposes where \coolname{}'s overhead comes from and what
would move the numbers.

\subsubsection{Decomposing the Overhead}

Runtime can be factored exactly into the executed
instruction count and the average cycles per instruction:
\[
  \frac{T_\text{sys}}{T_\text{native}}
  \;=\;
  \underbrace{\frac{I_\text{sys}}{I_\text{native}}}_{\text{instruction inflation}}
  \;\times\;
  \underbrace{\frac{\text{CPI}_\text{sys}}{\text{CPI}_\text{native}}}_{\text{per-instruction ratio}}.
\]

Figure~\ref{fig:runtime-instructions}(b) and
Table~\ref{tab:gmean-instructions} report the first term.
\coolname{} inflates the instruction stream by $7.12\times$
at \texttt{-O2} and $7.05\times$ at \texttt{-O3}, with a
tight per-benchmark range of $5.04\times$--$8.79\times$
(O2) and $4.78\times$--$9.05\times$ (O3). QEMU averages
$9.36\times$ and $9.56\times$ but spans nearly an order of
magnitude across benchmarks ($4.91\times$--$33.03\times$ at
\texttt{-O2}, $4.95\times$--$33.36\times$ at \texttt{-O3}).
Box64 inflates by $1.94\times$ and $1.96\times$.

\begin{figure*}[t]
  \centering
  \includegraphics[width=\linewidth]{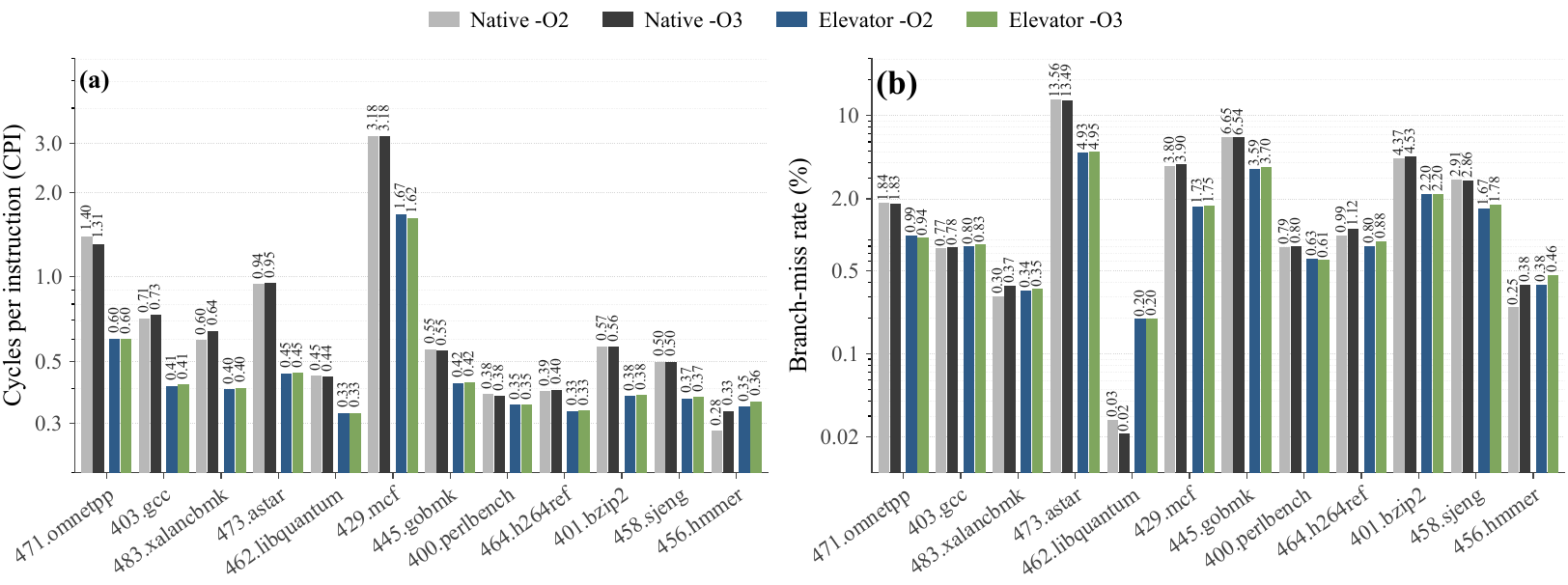}
  \caption{Microarchitectural behavior of \coolname{}
           against native AArch64 on SPECint~2006 at
           \texttt{-O2} and \texttt{-O3}: (a)~cycles per
           instruction (CPI); (b)~branch-miss rate. Log y-axes absorb the \texttt{429.mcf} (CPI) and \texttt{473.astar}
           (branch-miss) outliers. QEMU and Box64 are
           omitted because their per-instruction rates are
           diluted by translator-internal instructions and
           are not directly comparable on a per-instruction
           basis.}
  \label{fig:microarch}
\end{figure*}

Figure~\ref{fig:microarch}(a) reports the second term.
\coolname{}'s geometric-mean CPI is $0.443$ at \texttt{-O2}
and $0.445$ at \texttt{-O3}, against a native baseline of
$0.645$ and $0.655$; the CPI ratio is therefore
$0.69\times$ and $0.68\times$. Combining the two factors,
\coolname{}'s predicted slowdown is
$7.05 \times 0.68 \approx 4.8\times$, which is consistent with
the $4.79\times$ observed at \texttt{-O3}. Applying the same
decomposition to QEMU gives
$9.56 \times 0.81 \approx 7.7\times$. Both factors
contribute, but for \coolname{} the per-instruction ratio is
below unity and partially offsets the inflation --- meaning
\textbf{the overhead \coolname{} incurs relative to native
sits almost entirely in the size of the translated
instruction stream, not in how quickly that stream retires
on the target pipeline}.

\subsubsection{Where the Instructions Come From}
\label{sec:runtime-inflation}
The $7\times$ instruction inflation is produced by a small
number of structural sources that every x86--AArch64
translator must address. First, x86 computes six condition
flags (\textsf{PF}, \textsf{AF}, \textsf{SF}, \textsf{ZF},
\textsf{CF}, \textsf{OF}) on most arithmetic operations,
while AArch64 provides only the four NZCV flags natively;
faithful emulation of PF, AF, and the x86-specific
interactions with shifts and rotates requires multi-instruction sequences for every flag-writing x86 op. Second, x86's complex addressing modes (e.g.\ \texttt{[base + index*scale + disp]}) decompose into a short address-computation sequence in AArch64 whenever
the source operand cannot be expressed directly. Third, the AArch64 footprint of a translated x86 instruction depends on its semantics. Constructs such as \texttt{REP}-prefixed string operations and implicit partial-register updates sit at the heavier end of this spectrum, consistently producing more target instructions than their plainer counterparts. The tightness of \coolname{}'s inflation range ($5.04\times$--$8.79\times$) across the suite indicates that these sources scale roughly proportionally with the source instruction count rather than with any particular program pattern, which makes them a good target for per-instruction optimization.

\subsubsection{Microarchitectural Behavior}
\label{sec:runtime-microarch}

Figure~\ref{fig:microarch} reports per-benchmark CPI and branch-miss rate for \coolname{} and native. CPI is below native on 11 of 12 benchmarks at each optimization level, including the memory-bound \texttt{429.mcf} whose native CPI of $\approx 3.18$ drops to $\approx 1.67$ under translation. We interpret this not as a performance win but as a corollary of the translation: lowering x86 CISC sequences to AArch64 produces regular streams of simple, independent $\mu$ops that the out-of-order back-end can schedule in parallel more easily than the denser native code, even when the total amount of work is higher. The single exception is \texttt{456.hmmer}, whose hot kernel is already a tight native AArch64 loop with CPI below $0.34$ and little remaining headroom; the translated stream of simple $\mu$ops, while wider, retires slightly less densely (CPI $0.345$ at \texttt{-O2}, $0.360$ at \texttt{-O3}).

Branch and cache behavior is broadly preserved across translation. \coolname{}'s geometric-mean branch-miss rate is $1.01\%$ at \texttt{-O2} and $1.04\%$ at \texttt{-O3}, compared with $1.21\%$ and $1.27\%$ native; the rate is at or below native on 8 of 12 benchmarks at \texttt{-O2} and 9 of 12 at \texttt{-O3}. The handful of benchmarks whose \coolname{} branch miss rate marginally exceeds native (\texttt{403.gcc}, \texttt{456.hmmer}, \texttt{462.libquantum}, and at \texttt{-O2} also \texttt{483.xalancbmk}) all sit at very low absolute rates below $1\%$, where the percentage comparison is dominated by small absolute counts rather than by a systematic predictor degradation.


The conclusion from this subsection is negative. The pipeline, the branch predictor, and the cache hierarchy are not where \coolname{}'s overhead lives; it lives in the length of the translated instruction stream. That stream is determined at translation time, so \coolname{}'s cost on any input is a property of the shipped binary rather than its execution history, and any optimization that does not shorten that stream is unlikely to change the headline numbers.

\subsubsection{Predictability}
\label{sec:runtime-predictability}

The translated instruction stream is fixed at translation time. No translator state, code cache, or dispatch machinery runs alongside the translated binary, and no input can cause the stream to grow, shift, or respecialize. \coolname{}'s runtime cost on a given input is therefore a static property of the translated artifact: determined at translation time, inspectable without execution, and bounded above by whatever qualification inputs have already demonstrated.

This property is one only a fully static, heuristic-free translator can deliver. The per-benchmark spread in Figure~\ref{fig:runtime-instructions}(a) is its empirical shadow: QEMU's first-encounter translation, dispatch, and code-cache management work is part of the measured runtime, paid whenever an input drives execution into previously-unseen code. Box64 compresses this spread substantially, but no amount of engineering can drive the first-encounter cost to zero.

\subsection{Code Size Expansion}
\label{sec:code-size}

\begin{figure*}[h]
  \centering
  \includegraphics[width=\linewidth]{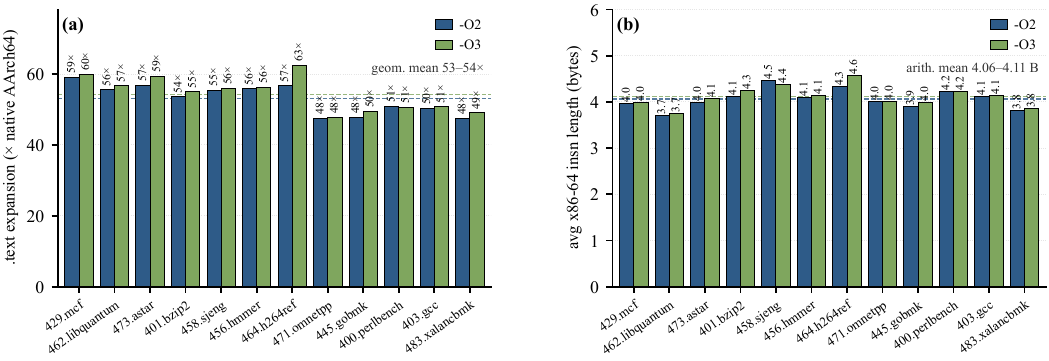}
  \caption{Code-size cost of \coolname{}'s superset translation on
           SPECint~2006 at \texttt{-O2} and \texttt{-O3}. (a)~Translated
           \texttt{.text} expansion relative to natively-compiled
           AArch64. (b)~Average x86-64 instruction length measured on
           each source binary.}
  \label{fig:code-size-expansion}
\end{figure*}

\begin{table*}[t]
  \centering
  \small
  \caption{Multiplicative decomposition of \coolname{}'s
           geometric-mean \texttt{.text} expansion. Product
           matches the gmean in
           Figure~\ref{fig:code-size-expansion}(a).}
  \label{tab:code-size-decomp}
  \begin{tabular}{@{}lp{0.45\linewidth}cc@{}}
    \toprule
    \textbf{Factor} & \textbf{Meaning} & \textbf{-O2} & \textbf{-O3} \\
    \midrule
    Per-instruction lowering
      & \coolname{}'s tile for a compiler-emitted x86
        instruction averages $\sim$7 AArch64 instructions
        (Section~\ref{sec:runtime-inflation}); native
        AArch64 compilation uses essentially the same
        instruction count as x86-64 (ratio $0.81$--$1.08$,
        gmean $0.96$), so this factor is attributable to
        the tile set rather than to cross-ISA structural
        differences.
      & $7\times$ & $7\times$ \\[3pt]
    Per-valid-byte candidate density
      & Tiles are emitted at every valid x86-64 decode
        position. Measured valid-decode rate is $\sim$91\%;
        combined with the $4.06$~B average x86-64
        instruction length
        (Figure~\ref{fig:code-size-expansion}(b)), this
        yields the number of tiles per real x86 instruction.
      & $3.71\times$ & $3.76\times$ \\[3pt]
    Non-real-offset encoding amplification
      & Only 1 of the 3.71 tiles per x86 instruction sits
        at the real instruction's byte offset; the other
        2.71 sit at mid-instruction offsets, where the
        decoder produces more complex x86 forms (prefixes,
        larger immediates, complex addressing modes) than
        compilers typically emit. Their tiles therefore
        require more AArch64 instructions than a
        real-instruction tile, raising the average tile size
        by this factor.
      & $2.04\times$ & $2.06\times$ \\
    \midrule
    \textbf{Product} & & $\mathbf{53.0\times}$ & $\mathbf{54.3\times}$ \\
    \bottomrule
  \end{tabular}
\end{table*}

\coolname{} emits an AArch64 sequence at every valid
source-byte offset of the x86-64 \texttt{.text}, and no
post-translation size reduction is applied. This follows
from the assumption-free stance the paper has taken
throughout \cite{bauman2018superset}. Every candidate-reduction path we are aware of (CFG-directed pruning,
probabilistic disassembly \cite{miller2019probabilistic})
either requires ground-truth information that the target
setting of stripped legacy binaries does not supply, or
introduces heuristics that can fail to translate x86
instructions. Optimizing the footprint is a choice we
defer to deployments where such information is present, or where the static binary size itself is a hard constraint. This subsection reports what that deferral costs.

\coolname{}'s translated \texttt{.text} is $47.5\times$ to $62.5\times$ larger than natively-compiled AArch64 \texttt{.text} across SPECint~2006 (Figure~\ref{fig:code-size-expansion}(a)), with geometric means of $53\times$ at \texttt{-O2} and $54\times$ at \texttt{-O3}: for every AArch64 instruction the native compiler emits, \coolname{} produces roughly $53$. The $7\times$ per-instruction lowering measured in Section~\ref{sec:runtime-inflation} accounts for only part of this cost; the remaining $\approx 7.5\times$ follows from superset translation. \coolname{} emits a tile for every valid byte offset, yielding roughly $3.71$ tiles per real x86 instruction given the suite's average $4.06$-byte encoding length (Figure~\ref{fig:code-size-expansion}(b)) and a measured valid-decode rate of $\approx 91\%$. The average tile is itself roughly twice the size of a real-instruction tile, because decodes starting at non-real offsets land on more complex x86 operations than
compilers typically emit. Table~\ref{tab:code-size-decomp} summarizes the three factors, whose product $7 \times 3.71 \times 2.04 \approx 53$ recovers the gmean.

Across the suite the expansion range is bounded. The three lowest-bloat benchmarks (\texttt{471.omnetpp}, \texttt{483.xalancbmk}, \texttt{445.gobmk}) use shorter x86-64 encodings than the suite mean on both density and amplification axes; the two highest-bloat benchmarks switch with opt level, with \texttt{464.h264ref} at \texttt{-O3}
($62.5\times$) driven by vectorized long-encoding hot paths and \texttt{429.mcf} at \texttt{-O2} ($59.2\times$) driven by a high amplification factor on a very small \texttt{.text}. Optimization level is otherwise immaterial: the geometric means differ by $\approx 3\%$ and all per-benchmark swings from \texttt{-O2} to \texttt{-O3} are within $\pm 3\times$ except \texttt{464.h264ref} ($+5.8\times$). All expansion ratios reported in this section are for \texttt{.text}. The overall binary footprint grows by a
smaller factor on benchmarks that bundle substantial static data, since \coolname{} passes data sections through unchanged (\texttt{445.gobmk} is the clearest case, carrying a $\sim\!4$\,MB pattern database alongside its $623$\,KB \texttt{.text}). The three decomposition factors admit independent optimization paths discussed together with the runtime-side options in Section~\ref{sec:runtime-improve}.

\subsection{Design Choices and Potential Optimizations}
\label{sec:runtime-improve}
Since the instruction stream dominates overhead, every
natural avenue for improving \coolname{} comes down to
emitting fewer instructions per translated x86 instruction.
Flag computation is the clearest case. \coolname{} currently
runs a backward flag-liveness pass over linear instruction
chains, which handles the common chained-arithmetic pattern
inside a single block. The flag computation work could be
reduced further by making the analysis per-flag rather than
per-EFLAGS. An x86 arithmetic instruction writes up to six
condition flags (PF, AF, SF, ZF, CF, OF), but the branch or
condition that eventually reads them almost always reads
only one or two, most often ZF for equality tests.
Computing and materializing only the subset of flags that is
actually live at each flag-writing site, rather than all
six, would shrink the code generated for every live flag
write.

A second potential optimization lies at the boundary between
the translated binary and external libraries. An x64 call
into a shared library expects arguments laid out in the x64
System~V convention (Figure~\ref{fig:exit-abi-translation});
AArch64 libraries expect the AAPCS64 convention. \coolname{}
bridges the two by conservatively copying up to $n$
potential stack argument slots from x64 positions to AArch64
positions on every external call, because the translator
cannot determine statically how many arguments the callee
consumes, nor what size and type each one has. The size
question matters as much as the count: both ABIs classify
each argument into register or stack slots based on its type
and byte size (scalars, small composites, and larger-than-16
-byte aggregates are handled differently), so even
recovering argument counts through heuristic dataflow
analysis is not sufficient to elide the copy. Resolving
callee signatures at translation time, from the dynamic
symbol table, library headers, or debug information, would
provide both pieces and allow the copy to shrink to the
actual number of argument bytes that the callee expects on
the stack. We leave this as future work, since \coolname{}
is designed to handle stripped legacy binaries where such
signature information is typically unavailable.

The larger determinant of \coolname{}'s instruction count is
how the tiles (Section~\ref{sec:tile-bank}) themselves are
written. \coolname{}'s code generator is deliberately
architecture-agnostic: it emits portable C that faithfully
encodes the semantics of each x86 instruction, with no
architecture-specific intrinsics, no inline assembly, and no
backend-specific library calls. The host compiler is
responsible for lowering the resulting C to target machine
code. For instance, specializing the vector-heavy tiles to emit AArch64 intrinsics directly would produce shorter sequences than the
compiler generates from portable C today, closing part of
the inflation gap to Box64.

We have not pursued that specialization for several
reasons. The tiles are the correctness-validated core of
\coolname{} (Section~\ref{sec:correctness}), and replacing
them with target-specific variants would reopen that
validation surface on a new architecture. The portable form
also means that support for new x86 extensions lands on
every backend simultaneously, without per-target engineering
work. \coolname{} itself is purely static and follows ABI
conventions only at external-library boundaries, so the
correctness of the tile set directly determines whether the
translated binary behaves the same as the input. The
resulting tiles may not emit the most compact AArch64
sequences the hardware allows, but they are
correctness-tested against the original x86 semantics, and
they remain the natural default if a target-specialized
variant is later introduced as an optional overlay. The question this paper answers is whether superset disassembly
with tile-based lowering produces a binary translator that
is correct, delivers microarchitectural behavior close to
native, and is competitive with a mature dynamic translator
at the whole-program level. The data in this section
supports each of those claims. Measuring the same benchmarks
under an AArch64-specialized tile set, or retargeting the
tiles to another ISA, is useful follow-up work rather than
evidence for or against the approach itself. 
Finally, given that \coolname{} already inflates binary size by a significant amount, adding specialized instruction sequences for hot code paths to achieve better performance is also a worthwhile direction for future optimizations.



\section{Current Implementation Limitations}

\textbf{ABI Differences.} Our zero-assumption argument-reorganizing ABI translation falls short for rare fundamental incompatibilities between source and target, such as structure-layout or argument-passing-convention mismatches. Identifying these and writing the interceptor functions that bridge them remains manual.


\textbf{Multi-Threaded Binaries.} \coolname{} currently supports only single-threaded binaries. The framework is largely designed to accommodate multi-threading, but two challenges remain. First, the \texttt{tcbhead\_t} structure that underlies thread-local storage (TLS) differs significantly between x64 and AArch64. Second, x64 uses a stronger (TSO) memory model than AArch64, and optimal fence placement to recover x86 ordering is undecidable at the binary level~\cite{deshpande2024polynima}; conservative placement degrades performance~\cite{AtoMig}. Hardware supporting the RCpc memory model (AArch64 v8.3~\cite{arm_armv8_3}) addresses this mismatch directly.


\textbf{Exception Handling.} \coolname{} does not yet support binaries that use exception handling, which primarily affects C++ exceptions (as in benchmark \texttt{471.omnetpp}). The main technical requirement is a stack unwinder that fetches x64 return addresses from the stack, since x64 stores them on the stack while AArch64 places them in a register. We have not implemented this because the engineering effort would not yield proportional scientific insight; this is a current implementation restriction, not a fundamental limitation of our approach.

\textbf{x64 Extensions.}
\coolname{} supports a substantial portion of x64’s instruction set, but several extensions remain unsupported. While supporting the remainder of SSE would be straightforward, AVX2 and later expand the existing 128-bit registers to 256-bit and 512-bit widths. Both exceed AArch64’s SIMD register width of 128-bits and would require implementing an additional memory-backed register context, or using the AArch64’s SVE extensions.

\textbf{Self Modifying and JIT-Compiled Code.} \coolname{}, like all fully static binary rewriters, does not support self modifying or just-in-time-compiled code.

\section{Future Work}
Key priorities for extending this work to enhance its scope and applicability include support for multi-threaded input binaries and expansion to additional target instruction set architectures. Beyond these architectural expansions, we plan to implement several optimization strategies, including dead code elimination and optimized flag computation to reduce both binary size overhead and runtime performance impact.

\section{Summary and Conclusion}

We have presented what we believe is the first fully static whole-program binary cross-compiler from one ISA to another. Previous static translators have all relied on heuristics; such approaches become less and less practical as the size of the input program grows, since successful translation depends on successive heuristics getting \emph{every} decision right. Our approach doesn’t use any heuristics at all and hence is practical for input programs of any size and complexity from any toolchain; our prototype implementation is mature enough to handle the entire SPECint 2006 benchmark suite, enabling a realistic evaluation on a range of input programs closely resembling real-world requirements. We can also translate input programs containing exotic overlapping/nested/obfuscated code constructs that existing binary translators have not been able to handle correctly. Performance-wise, our current, not yet fully optimized implementation already matches or outperforms the state-of-the-art QEMU JIT-accelerated emulation framework on a majority of SPECint~2006 benchmarks.

From a practical perspective, our approach lowers the risk of deploying cross-ISA translation since the code that will ultimately be executed is generated in its entirety ahead of time. It can therefore be rigorously tested, certified, and possibly cryptographically signed in the same manner as traditional native binaries. In contrast, approaches based on emulation and JIT compilation implicitly depend on additional runtime components and any tests validated under any specific version of these runtime components don’t necessarily transfer to any other version of these same components.

\section{Acknowledgments}
We express our gratitude to Chinmay Deshpande, Fabian Parzefall, David Gens, Mitchel Dickerson, and Ryan Snyder for their feedback.

This research was funded in part by the Defense Advanced Research Projects Agency under contracts 140D04-23-C-0070. Any opinions, findings, conclusions, or recommendations expressed in this material are those of the author(s) and do not necessarily reflect the views of the Defense Advanced Research Projects Agency (DARPA).


\bibliographystyle{ACM-Reference-Format}
\bibliography{reference}

\end{document}